# Crossover from Conventional to Unconventional Superconductivity in 2M-WS$_2$

*Piumi Samarawickrama[1,2], Joseph McBride[3], Sabin Gautam[1,2], ZhuangEn Fu[1,2], Kenji Watanabe[4], Takashi Taniguchi[5], Wenyong Wang[1,2], Jinke Tang[1,2], John Ackerman[6], Brian M. Leonard[3,2] and Jifa Tian[1,2*]*

1.   Department of Physics and Astronomy, University of Wyoming, Laramie, Wyoming 82071, USA

2.   Center for Quantum Information Science and Engineering, University of Wyoming, Laramie, Wyoming 82071, USA

3.   Department of Chemistry, University of Wyoming, Laramie, Wyoming 82071, USA.

4.   Research Center for Electronic and Optical Materials, National Institute for Materials Science, 1-1 Namiki, Tsukuba 305-0044, Japan

5.   Research Center for Materials Nanoarchitectonics, National Institute for Materials Science, 1-1 Namiki, Tsukuba 305-0044, Japan

6.   Department of Chemical and Biomedical Engineering, University of Wyoming, Laramie, Wyoming 82071, USA

* Email: jtian@uwyo.edu

ABSTRACT

Leveraging reciprocal-space proximity effect between superconducting bulk and topological surface states (TSSs) offers a promising way to topological superconductivity. However, elucidating the mutual influence of bulk and TSSs on topological superconductivity remains a challenge. Here, we report pioneering transport evidence of a thickness-dependent transition from conventional to unconventional superconductivity in 2M-phase WS$_2$ (2M-WS$_2$). As the sample thickness reduces, we see clear changes in key superconducting metrics, including critical temperature, critical current, and carrier density. Notably, while thick 2M-WS$_2$ samples show conventional superconductivity, with an in-plane (IP) upper critical field constrained by the Pauli limit, samples under 20 nm exhibit a pronounced IP critical field enhancement, inversely correlated with 2D carrier density. This marks a distinct crossover to unconventional superconductivity with strong spin-orbit-parity coupling. Our findings underscore the crucial role of sample thickness in accessing topological states in 2D topological superconductors, offering pivotal insights into future studies of topological superconductivity.

The quest for topological quantum states in superconductors has attracted significant attention in condensed matter physics, fueled by the promise of uncovering novel phenomena and potential applications in quantum computing. [1, 2] The two-dimensional (2D) van der Waals (vdW) superconductors, [3] epitomized by thin, crystalline atomic layers and rich types of crystal structure and symmetry, offer an unparalleled platform for exploring and manipulating exotic superconducting and topological quantum states. In particular, the reduced dimensionality of 2D vdW superconductors amplifies quantum effects, creating an ideal setting for investigating the interplay between superconductivity, dimensionality, spin-orbit coupling (SOC), and topological surface states (TSSs). This interplay leads to exotic phenomena and physical properties, such as Ising superconductivity[4, 5], topological band inversion with spin-orbit parity coupling (SOPC) [6], and topological superconductivity[7-10]. For instance, Ising superconductivity, a phenomenon found in specific 2D superconductors, manifests in two types: type I, observed in superconductors lacking inversion symmetry (such as gated $MoS_2$[11] and $WS_2$[12] and monolayer $NbSe_2$[13]) and type II, present in superconductors with both a centrosymmetric structure and rotational symmetry[5]. Both types of Ising superconductors can lead to unconventional superconducting states and enhance the materials' resilience to in-plane (IP) magnetic fields. Theorists also predicted that in superconductors (such as $1T$'-$WTe_2$) with centrosymmetric structures but not certain rotational symmetry, SOPC near the band inversion point could account for the enhanced upper critical field. [6] Furthermore, the intriguing combination of $s$-wave superconductivity with TSS provides a promising way for realizing topological superconductivity[14, 15]. Inherently governed by topology, topological superconductors (TSCs) can host Majorana zero modes (MZMs) - quasiparticles crucial for achieving fault-tolerant quantum computing[15, 16]. However, the intricate interplay between bulk and TSSs in engineering topological superconductivity in 2D superconductors realized by the real or reciprocal-space proximity effect remains a significant, yet-to-be-resolved challenge.

2M phase $WS_2$ (2M-$WS_2$), a recently recognized TSC candidate via the reciprocal-space proximity effect[17], has been drawing significant attention in the scientific community[18-28]. The 2M-$WS_2$ has a centrosymmetric structure with $C_{2/m}$ symmetry, while its monolayers display a $1T'$ phase by $W$-$W$ zigzag chains along the 'a' axis,[17] as depicted in Fig. 1a. These monolayers are stacked along the 'c' axis through a translational operation. The material has sparked further interest due to its TSS and the potential for hosting MZMs[20, 23-25] in the vortex cores. Notably, recent studies have demonstrated that bulk states in 2M-$WS_2$ feature single-gap $s$-wave superconductivity[19,21 23-25] and a zero-bias conductance peak at the cores of magnetic vortices on the surface of 2M-$WS_2$[20] with an unusual enhancement of the upper critical field in its atomically thin layers[26]. Despite the promising advances among the 2D TSC candidates created by the reciprocal-space proximity effect, the crucial role of thickness in modulating the proximity-induced surface (topological or unconventional) superconductivity and their bulk superconductivity is still elusive. Nevertheless, the 2M-WS$_2$, with a simple composition and layered structure, represents an ideal platform to address these challenges.

In this study, we explore the superconductivity in 2M-WS$_2$ thin layers through electrical transport measurements, focusing on how the TSS affects its unconventional superconductivity in the clean limit. Our samples, ranging in thickness from ~ 3 nm (~5 layers) to 50 nm (~83 layers), were prepared using the scotch tape method[28]. We first present the systematic studies on the layer dependences of the samples' corresponding critical temperature, critical current, and carrier density. We then uncover a pronounced anisotropy and 2D nature of superconductivity in 2M-WS$_2$ thin layers. We further reveal a non-trivial transition from conventional to unconventional superconductivity as the sample thickness decreases below 20 nm. The unconventional superconductivity of 2M-WS$_2$ layers is marked by a significant enhancement of the IP upper critical field, inversely correlated with its 2D carrier density, showing the transport evidence of the existence of spin-orbit parity coupled superconductivity. Our study unveils a pioneering advance in understanding the pivotal influence of sample thickness on the transport properties of 2M-WS$_2$, highlighting its critical role in modulating both the conventional bulk and the proximity-induced surface superconductivity in 2D superconductors.

Like other 2D materials, the sample thickness was determined using atomic force microscopy (AFM) and optical contrast. Figure 1b shows an AFM image of a 2M-WS$_2$ (3 nm) flake on a SiO$_2$/Si substrate. To confirm the 2M phase, Raman spectroscopy measurements (right panel of Fig. 1c) with a 532 nm laser were performed on a 2M-WS$_2$ sample with different thicknesses (left panel of Fig. 1c). Locations marked "1" to "5" on the flake indicate where the corresponding Raman analysis was performed. We find that the intensities for all the Raman peaks generally increase with the decreasing layer thickness, without any noticeable shift in peak positions for samples thicker than 3 nm. Given the metastable nature of the 2M-WS$_2$[27], to study the intrinsic transport properties of 2M-WS$_2$ thin layers, we directly transferred the few-layer 2M-WS$_2$ onto the pre-made Hall-bar electrodes, avoiding any potential contamination and phase change during the standard nanodevice fabrication procedure. An optical image of a Hall-bar device for the 3-nm-thick layer is shown in the inset of Fig. 1d, along with the corresponding temperature dependence of longitudinal resistance ($R_{XX}$). A clear metallic to superconducting transition is observed at a transition temperature ($T_C$) of 6.98 K. In this work, we define $T_C$ as the temperature where the sample resistance falls to 75% of its normal value $R_N$. The narrow temperature range of approximately 0.5 K for the metal-to-superconductor transition (Inset of Fig. 1d) with a residual resistivity ratio (RRR) of ~ 11 for the 3 nm-thick sample demonstrates the high quality of the prepared 2M-WS$_2$ atomic layers.

To study how thickness affects the transport properties of 2M-WS$_2$, we conducted measurements on samples with thicknesses ranging from 3 to 50 nm. We note that the individual 2M-WS$_2$ layer with a uniform thickness (roughness around or less than 1 nm, Fig. S1) has been selected for device fabrication. We first identified the carrier density and mobility via Hall effect measurements. The representative results of the Hall resistance ($R_{XY}$) results are shown in Fig. S2. We note that all the Hall resistance curves in Fig. S2b-g have been subjected to symmetrical correction (Fig. S2a). Since the $R_{XY}$-$B$ plots are mostly linear, we employed the single-band model to fit our data and extracted the corresponding carrier density and Hall mobility. The extracted results for different 2M-WS$_2$ layers at different temperatures are summarized in Figs. 2a and S3a. We see that the carrier density for 2M-WS$_2$ generally decreases as both the temperature and thickness decrease. For instance, the 2D carrier density reduces more than one order of magnitude as the sample thickness decreases from 40 nm (~ $4.8 \times 10^{15} cm^{-2}$) to 3 nm (~ $4.2 \times 10^{14} cm^{-2}$) at 10 K, indicating the reduced contribution of bulk states to the transport. In contrast, the Hall mobility shows an opposite temperature dependence compared to that of carrier density (Fig. S3a). Despite some variations, the mobility generally decreases as sample thickness decreases. We further calculated the mean free path (Fig. S3b) $l = hk_F/2\rho_0 Ne^2$ of the samples, where $k_F$ ($= (3\pi^2 N)^{1/3}$) is the Fermi wave number, $N$ is the 3D carrier density, and $\rho_0$ is the resistivity prior to the superconducting transition. We see that the mean free path shares similar temperature and thickness dependences as those of mobility.

Figure 2b shows the normalized sheet resistance ($\frac{R_S}{R_N}$) as a function of temperature for various 2M-WS$_2$ thin flakes. Figure 2c summarizes the thickness dependence of $T_C$ of the 2M-WS$_2$ layers, where $T_C$ is reduced from 8.76 to 6.98 K as the thickness decreases from 40 nm to 3 nm. We note that the thickness dependence of $T_C$ is much weaker compared with other 2D superconductors, such as NbSe$_2$[13]. To further characterize the quality of the prepared 2M-WS$_2$ thin flakes, we calculated the corresponding $RRR$ ($= R_{RT}/R_N$). It is known that a higher value of $RRR$ indicates fewer defects and impurities scattering in a material, suggesting that the transport properties are predominantly governed by phonons and reflecting the material's intrinsic characteristics. We observe that the $RRR$ decreases from 103 to 11 as the sample thickness decreases from 40 to 3 nm (inset of Fig. 2c). Despite this reduction, the values remain high, confirming the high quality of our samples. We further measured the $I$-$V$ characteristics of 2M-WS$_2$ layers at different temperatures (Fig. 2d and Fig. S4). The inset of Fig. 2d shows the $I$-$V$ traces of a 15 nm-thick 2M-WS$_2$ flake. Figure 2d summarizes the thickness dependence of the critical current for different 2M-WS$_2$ layers measured at temperatures 0.2 K lower than their corresponding $T_C$'s. As expected, the critical current again decreases with the decreasing sample thickness.

Figure 3a shows the $I$-$V$ traces of a 9 nm-thick sample as a function of temperature (plotted in log-log scale). We see that the $I$-$V$ characteristics follow a power law dependence ($V \sim I^\alpha$), consistent with Berezinskii-Kosterlitz-Thouless (BKT) transition[29] for 2D superconductors. The hallmark of the transition is the deviation from a linear relationship to a power-law dependence in the $I$-$V$ characteristics[26, 29], a phenomenon attributed to the movement of free vortices within the system[29-31]. This is specifically observed at $\alpha = 3$, as delineated by the black dotted lines in Fig. 3a and S5. Figure 3b summarizes the temperature dependence of $\alpha$ for 2M-WS$_2$ layers with different thicknesses. Initially, $\alpha$ shows a weak temperature dependence, followed by a sharp increase, surpassing the value of 3, where the superconductor undergoes a transition from 3D to 2D. We find that the BKT transition temperature ($T_{BKT}$) (defined as the critical temperature at $\alpha = 3$) decreases with the decreasing sample thickness, as shown in the inset of Fig. 3b. We see that $T_{BKT}$'s are only slightly smaller than their corresponding $T_C$'s (Fig. 2c), consistent with the behavior of other 2D superconductors[32].

We further performed the transport measurements under both IP and out-of-plane (OOP) magnetic fields (Figs. 3 and S6,7). Figure 3c shows the normalized sheet resistance ($R_S/R_N$) as a function of the OOP magnetic field measured at 1.5 K. We see that the upper critical field $B_{C2}$, defined as the magnetic field at which the $R_S/R_N$ drops to 75% of its value in the normal state, is weakly dependent on the sample thickness as summarized in the inset of Fig. 3c. Figures 3d and e show the representative temperature dependences of $R_S$ under OOP and IP magnetic fields, respectively. In both cases, there is a noticeable broadening of the superconducting transition region as the magnetic field increases. However, this broadening effect is significantly more pronounced under OOP magnetic fields, suggesting a marked anisotropy. Strikingly, superconductivity for the thin samples (Figs. 3e and S7) remains unsuppressed even under a 12 T IP magnetic field.

To further understand the observed anisotropy and 2D superconductivity, we analyzed the angular dependence of $B_{C2}$ for samples with different thicknesses (Figs. 4a and S8a,b). The representative result for a 6 nm-thick device is shown in Fig. 4a, where $\theta$ is the angle between the applied magnetic field and the sample plane. We observe a pronounced peak of $B_{C2}$ at $\theta = 0°$, which then decreases sharply as $\theta$ increases to 10°, before exhibiting a weak dependence on $\theta$. The angle dependence of the upper critical field is generally described either by the anisotropic 3D Ginzburg-Landau (GL) model, $\left(\frac{B_{C2}(\theta)\cos(\theta)}{B_{C2}^{\perp}}\right)^2 + \left(\frac{B_{C2}(\theta)\sin(\theta)}{B_{C2}^{\parallel}}\right)^2 = 1$ (green line) or 2D Tinkham's formula $\left|\frac{B_{C2}(\theta)\cos(\theta)}{B_{C2}^{\perp}}\right| + \left(\frac{B_{C2}(\theta)\sin(\theta)}{B_{C2}^{\parallel}}\right)^2 = 1$ (black line).[33] The 3D GL model explains the anisotropic nature of the upper critical field in 3D superconductors, whereas the Tinkham model is better suited for describing 2D systems. We see that our data can only be fitted by the 2D-Tinkham model, suggesting the 2D nature of the superconductivity in the measured 2M-WS$_2$ thin layers. We further calculated the IP and OOP coherence lengths for all the measured samples using the following two equations: $B_{C2}^{\perp} = \frac{\Phi_0}{2\pi\xi_{\parallel}^2(0)}$ and $B_{C2}^{\parallel} = \frac{\Phi_0}{2\pi\xi_{\perp}(0)\xi_{\parallel}(0)}$. Figure S8d summarizes the corresponding thickness dependences of the coherence length (left axis) and mean free path (right axis) of the 2M-WS$_2$ samples. Both the IP and OOP coherence lengths fall within a few nanometers range, while the mean free path spans a few micrometers. For example, a sample with a thickness of 3 nm has a mean free path of $2.73 \times 10^{-6}$ m and an IP coherence length of $8.35 \times 10^{-9}$ m. In the case of a 40 nm-thick flake, these values are $5.86 \times 10^{-6}$ m and $2.77 \times 10^{-9}$ m, respectively. The ratio between the mean free path and the coherence length is notably large, ranging from 300 to 2000 depending on the sample thickness, indicating that our samples are in the clean limit. We note that the OOP coherence lengths are around 1- 2 nm with a weak dependence on thickness, where the IP coherence length increases notably in samples thinner than 20 nm (Fig. S8d).

We further analyzed the temperature dependence of $R_{XX}$ for different samples under both the IP and OOP upper critical fields. Figure 4b summarizes the corresponding $B_{C2}$ normalized to the Pauli paramagnetic limit ($B_P = 1.84 T_C$), $B_{C2}/B_P$, as a function of $T_C$. Contrary to the previous report, [26] we see that both the normalized IP and OOP upper critical fields ($B_{C2}/B_P$) show a linear relationship with $T_C$. We employed the linearized GL theory, [33] $B_{C2}^{\perp} = \frac{\Phi_0}{2\pi \xi^2(0)} \left(1 - \frac{T}{T_C}\right)$, to fit the results for OOP fields, as shown in Fig. 4b, where $\Phi_0$ is the flux quanta, and $\xi$ is the GL coherence length. We find that the OOP upper critical fields at zero temperature, $B_{C2}^{\perp}(0)$, fall below $B_P$. Given the linearity of the $B_{C2}/B_P$ vs $T_C$ curves for the IP magnetic fields, the commonly used 2D GL theory for the IP magnetic fields, $B_{C2}^{\parallel} = \frac{\Phi_0 \sqrt{12}}{2\pi \xi(0) d} \left(1 - \frac{T}{T_C}\right)^{\frac{1}{2}}$, does not fit our data. This discrepancy is evident in the fittings (two dotted curves) for the samples with 43 & 6 nm thicknesses, as shown in Fig. 4b. Therefore, to estimate the IP upper critical field ($B_{C2}^{\parallel}(0)$) at zero temperature, we employed the linearized GL formula to fit our results, as shown in Fig. 4b. Interestingly, we find that for samples thicker than 20 nm, the $B_{C2}/B_P$ versus $T_C$ curves show a relatively weak temperature dependence and converge into a narrow band with a similar slope. The superconductivity is suppressed at the Pauli limit, indicating their conventional *s*-wave nature. Remarkably, when the sample thickness is below 20 nm, the $B_{C2}^{\parallel}(0)$ values undergo a pronounced transition from below $B_P$ to substantially above $B_P$. Furthermore, the slope of the $B_{C2}/B_P$ versus $T_C$ curve increases with the reduced sample thickness. These striking results indicate a non-trivial and thickness-induced crossover from conventional to unconventional superconductivity in 2M-WS$_2$, which has not been revealed before. Supported by the recent experimental and theoretical studies,[18-28] the anticipated surface topological superconductivity in 2M-WS$_2$ is attributed to the proximity effect between conventional *s*-wave superconductivity in the bulk states and the TSSs in the reciprocal space. Analogous to the competing behaviors between bulk and TSSs in 3D topological insulators, the interplay between bulk superconductivity and surface unconventional superconductivity in 2M-WS$_2$ is expected to significantly impact transport results. However, to date, no studies have directly addressed this phenomenon in the context of 2D topological superconductivity.

Next, we explore the nature of the unconventional superconductivity of samples thinner than 20 nm (Figs. 4b and S8c). Unconventional superconductors can feature an IP $B_{C2}$ exceeding the Pauli paramagnetic limit, attributing to various factors, including strong SOC[11-13], unconventional pairing mechanisms[34], or multi-band superconductivity[35], etc. Since 2M-WS$_2$ preserves inversion symmetry, conventional SOC attributed to inversion symmetry breaking cannot account for the significantly enhanced $B_{C2}^{\|}$. We also rule out multi-band superconductivity, quantum fluctuation induced by sample disorders, and finite-momentum Cooper pairs as the causes because no obvious changes in the band structure and/or sample quality are expected when the sample is thinner than 20 nm. We note that determining whether the orbital effect plays a role requires further experimental studies. We notice that an earlier theoretical work predicted that the SOPC arising from the coupling of spin, momentum and band parity could exist near the band inversion region of 1$T$'-WTe$_2$ with inversion symmetry[6]. Subsequently, it was suggested that the SOPC could exist near the topological band crossing of 2M-WS$_2$,[26] and this might be the driving force behind the enhancement of $B_{C2}^{\|}$. One of the hallmark signatures for superconductivity with SOPC is that $B_{C2}^{\|}$ shows a marked carrier density (Fermi level) dependence as the SOPC becomes prominent only near the band crossing points. However, this aspect has not been investigated before[26]. From Fig. 2a, we revealed a distinct thickness-dependent carrier density of 2M-WS$_2$, indicating that thickness might be utilized to modulate the expected SOPC in 2M-WS$_2$. Figure 4c summarizes the thickness dependences of the 2D carrier densities measured at 10 K and the slopes ($\frac{dB_{C2}^{\|}}{dT_C}$) of the $B_{C2}^{\|}$ versus $T_C$ curves (Fig. 4b). We see that the 2D carrier density decreases ~5 times as the sample thickness reduces from 50 to 6 nm, while the slope $\frac{dB_{C2}^{\|}}{dT_C}$ increases ~3 times. Our analysis reveals a pronounced correlation between the $B_{C2}^{\|}$ and 2D carrier density of 2M-WS$_2$. These findings suggest that SOPC underpins the enhanced $B_{C2}^{\|}$ in 2M-WS$_2$ thin layers (sub-20 nm) as the contribution from TSS is increased. Concurrently, the non-trivial crossover from conventional to spin-orbit-parity coupled superconductivity (Fig. 4b), aligns well with our observations. Thus, our work highlights the critical role of sample thickness in probing the topological properties of this type of material through electrical transport measurements, paving the way for future studies on thickness-driven hybridization of MZMs on the top and bottom vortices and 1D vortex-line topology.

In summary, our study has unveiled critical insights into the effect of sample thickness on the unconventional superconductivity in 2M-WS$_2$ thin layers. Through magnetotransport measurements, we have elucidated the 2D nature and pronounced anisotropy of the superconductivity in our samples. We demonstrated pioneering evidence of a thickness-dependent crossover from conventional to unconventional superconductivity in the 2M-WS$_2$ thin layers. We further identified the critical role of 2D carrier density in understanding the significant enhancement of the IP upper crucial field, indicating the existence of spin-orbit-parity coupled superconductivity. Our findings thus provide critical insights into the interplay between bulk and surface states, highlighting the complex mechanisms at play and underscoring the potential of 2M-WS$_2$ as a platform for studying topological superconductivity.

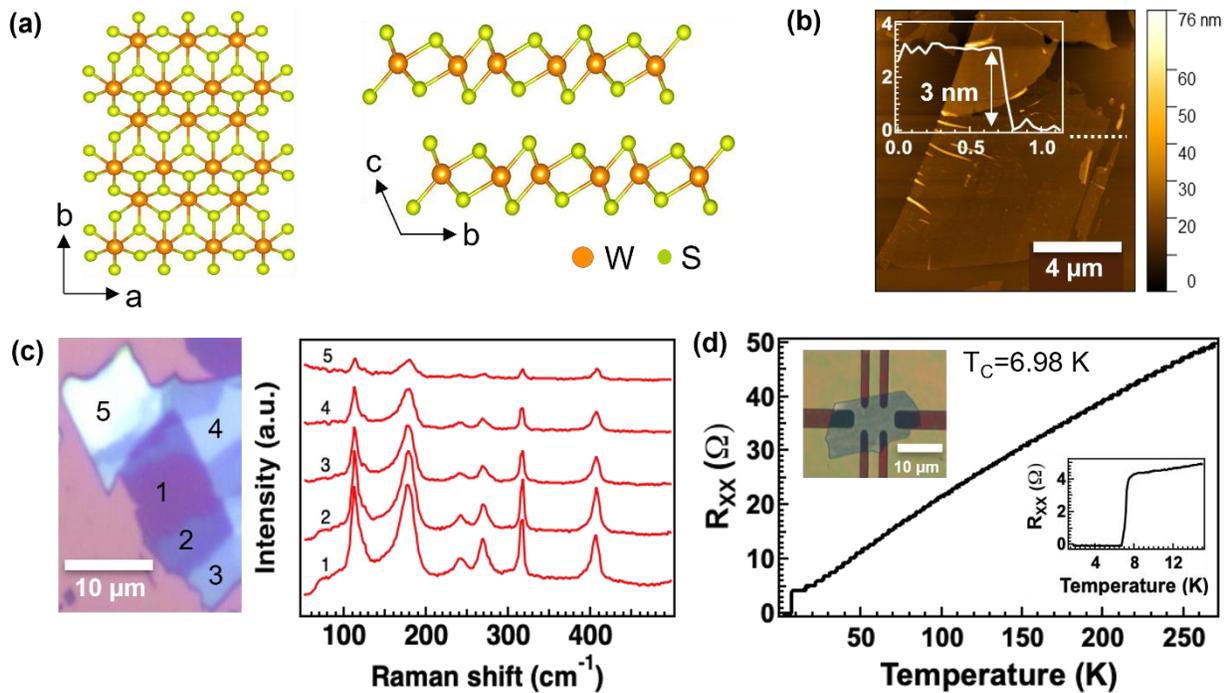

**Figure 1.** Raman spectroscopy and superconductivity of 2M-WS$_2$ thin layers. (a) Top and side view of the lattice structure of 2M-WS$_2$. (b) AFM image of a 3 nm-thick 2M-WS$_2$ flake. Inset: the thickness profile of the flake along the white dotted line. (c) Optical image (left) of a 2M-WS$_2$ flake with different thicknesses and Raman spectra (right) taken at different locations labeled with numbers "1" to "5" on the flake. (d) Longitudinal resistance ($R_{XX}$) versus temperature of a 3 nm-thick 2M-WS$_2$ device. Inset: zoomed-in view near the superconducting transition temperature.

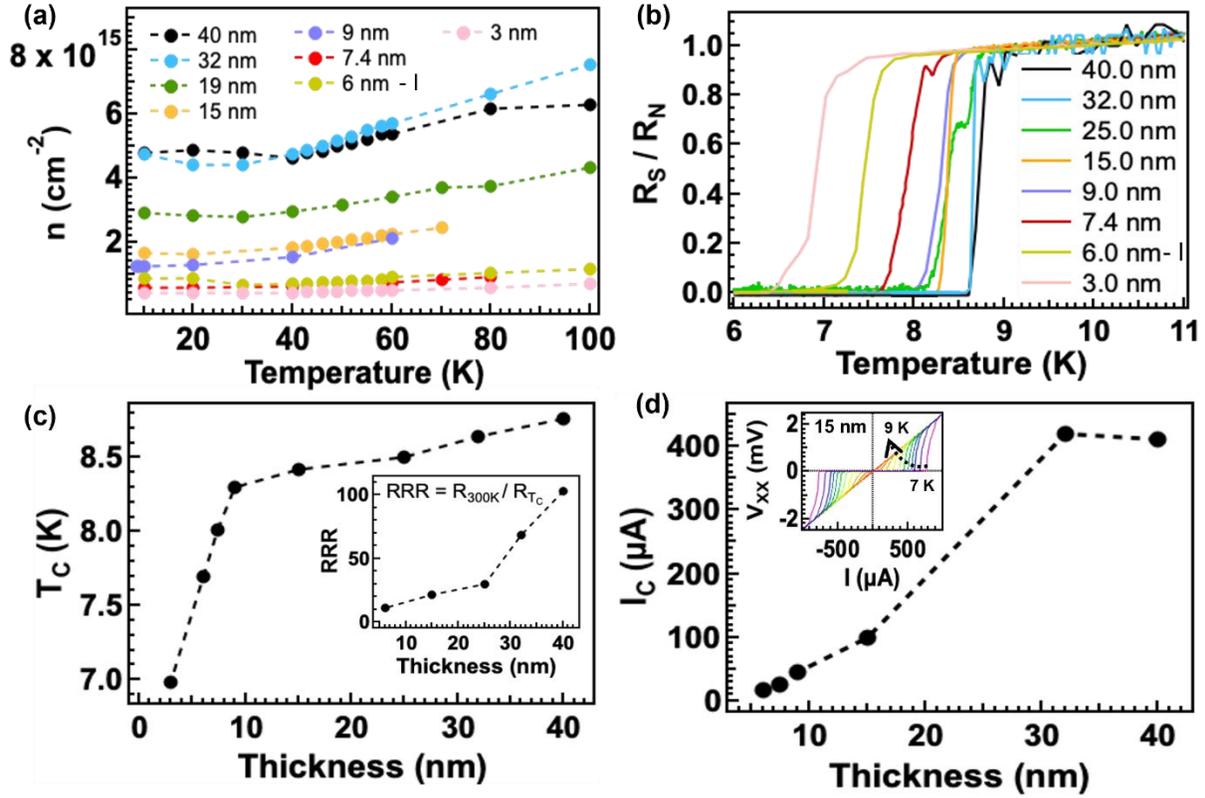

**Figure 2.** Thickness dependent superconductivity of 2M-WS$_2$. (a) Temperature dependence of the 2D carrier density of 2M-WS$_2$ flakes with various thicknesses. (b) The normalized sheet resistance as a function of temperature for 2M-WS$_2$ flakes. (c) Transition temperature ($T_C$) as a function of sample thickness. Inset: The corresponding thickness-dependent residual resistivity ratio (RRR) is calculated by dividing the sheet resistance at room temperature by the normal state resistance at the onset of the superconducting transition. (d) Critical current as a function of thickness measured at a temperature that is 0.2 K below the corresponding $T_C$. Inset: *I-V* characteristics of a 15 nm-thick device measured within a temperature range of 7 to 9 K.

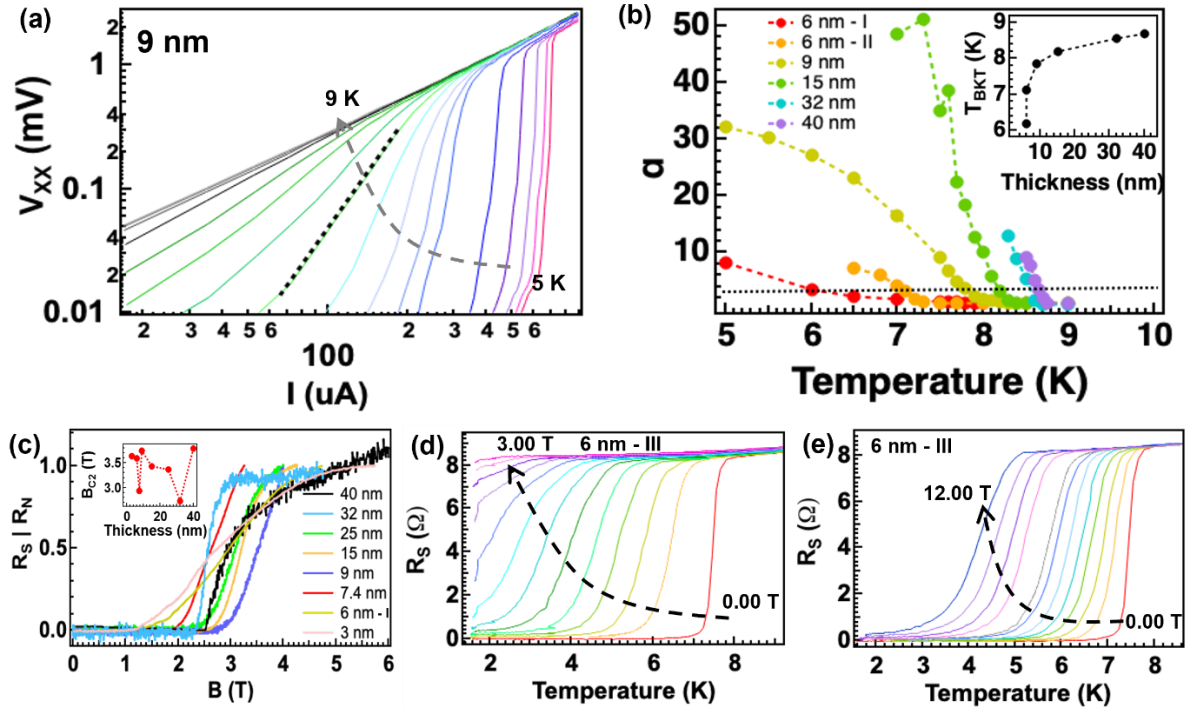

**Figure 3.** Berezinskii-Kosterlitz-Thouless transition and strong anisotropy in 2M-WS$_2$. (a) *I-V* relationship of a 9 nm-thick 2M-WS$_2$ device at various temperatures plotted in a log-log scale. The black dotted line represents the power law dependence $V \propto I^\alpha$ with $\alpha$ = 3. (b) The extracted $\alpha$ values as a function of temperature for different samples. Dotted line represents $\alpha$ = 3. The inset shows the extracted T$_{BKT}$ values for the corresponding samples. (c) Normalized sheet resistance as a function of the applied out-of-plane (OOP) magnetic fields for different 2M-WS$_2$ flakes measured at 1.5 K. The inset shows the thickness dependence of the extracted upper critical field. Temperature dependence of the sheet resistance for a 6 nm-thick device measured at different (d) OOP and (e) IP magnetic fields. The step size of the OOP magnetic field is 0.25 T and that for IP field is 1 T.

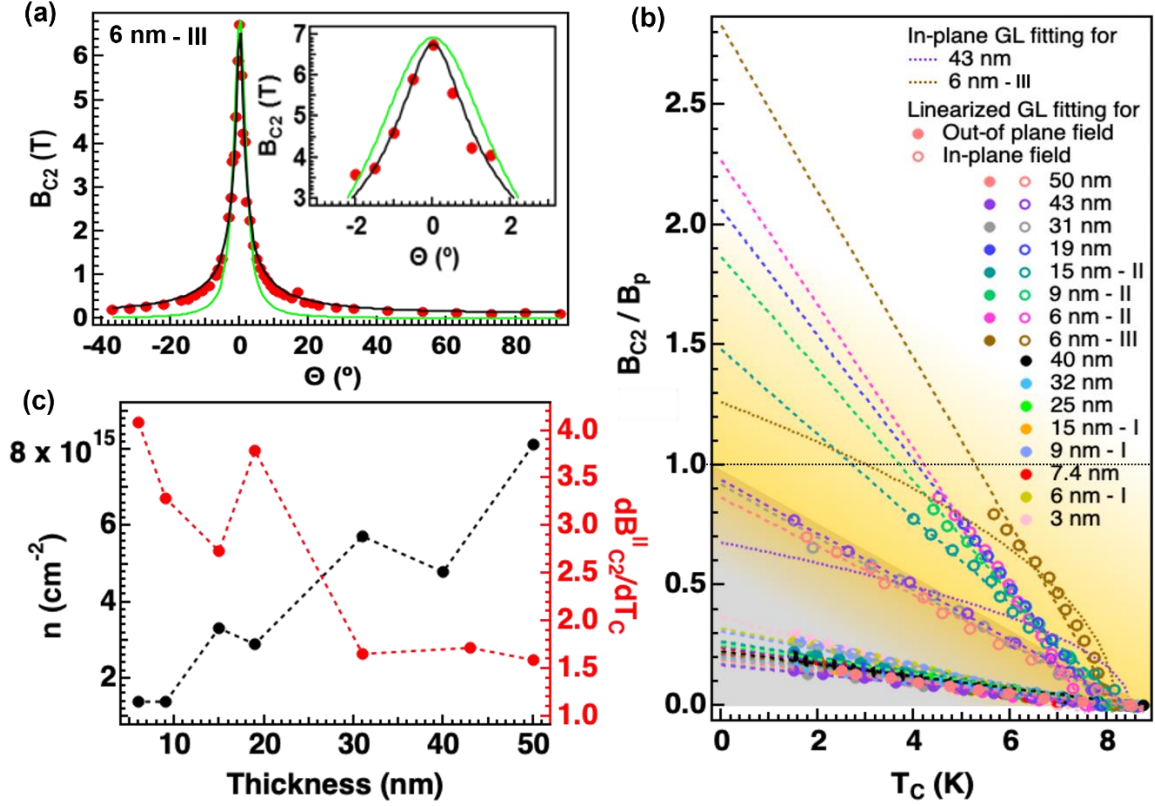

**Figure 4.** Thickness-dependent transition from conventional to SOPC superconductivity in 2M-$WS_2$. (a) Angular dependence of the upper critical field for a 6 nm flake measured at 7 K. The results are fitted by both 2D Tinkham model (black curve) and 3D anisotropic GL model (green curve). (b) The calculated $B_{C2}/B_P$ as a function of $T_C$. Black dotted line (horizontal) denotes the Pauli paramagnetic limit ($B_P$). Dashed lines are the fittings using the linearized GL formula, while dotted lines are the fittings of IP GL formula. (c) 2D carrier density (left axis) and $\frac{dB_{C2}^{\parallel}}{dT_C}$ (right axis) versus thickness for different samples.

**Corresponding Author**

Jifa Tian – Department of Physics and Astronomy and Center for Quantum Information Science and Technology, University of Wyoming, Laramie, Wyoming, 82070, USA; Email: e-mail: jtian@uwyo.edu

**Author Contributions**

J.T. conceived the project. P.S. fabricated the devices. P.S., S.G. and Z.F. performed the measurements. J.M., J.A. and B.M.L. grew the bulk 2M-$WS_2$ crystals. K.W. and T.T. grew the bulk h-BN crystals. P.S., W.W., J.Ta. and J.T. analyzed the data. P.S and J.T. wrote the manuscript. All authors discussed the results and commented on the manuscript.

**Notes**

The authors declare no competing financial interest.

ACKNOWLEDGMENT

This research was mainly supported by the U.S. Department of Energy, Office of Basic Energy Sciences, Division of Materials Sciences and Engineering under award DE-SC0021281 for sample and device fabrication and after its completion by DE-SC0024188 for transport measurements. J.T. also acknowledges the financial support of the U.S. National Science Foundation (NSF) grant 2228841 for data analysis, K.W. and T.T. acknowledge support from the JSPS KAKENHI (Grant Numbers 20H00354 and 23H02052) and World Premier International Research Center Initiative (WPI), MEXT, Japan.

REFERENCES

(1) Clarke, J.; Wilhelm, F. K. Superconducting Quantum Bits. *Nature* **2008**, *453* (7198), 1031-1042.

(2) Li, Y.; Xu, Z.-A. Exploring Topological Superconductivity in Topological Materials. *Adv. Quantum Technol.* **2019**, *2* (9), 1800112.

(3) Qiu, D.; Gong, C.; Wang, S.; Zhang, M.; Yang, C.; Wang, X.; Xiong, J. Recent Advances in 2D Superconductors. *Adv. Mater.* **2021**, *33*, 2006124.

(4) Wickramaratne, D.; Khmelevskyi, S.; Agterberg, D. F.; Mazin, I. I. Ising Superconductivity and Magnetism in $NbSe_2$. *Phys. Rev. X* **2020**, *10* (4), 041003.

(5) Wang, C.; Lian, B.; Guo, X.; Mao, J.; Zhang, Z.; Zhang, D.; Gu, B.-L.; Xu, Y.; Duan, W. Type-II Ising Superconductivity in Two-Dimensional Materials with Spin-Orbit Coupling. *Phys. Rev. Lett.* **2019**, *123* (12), 126402.

(6) Xie, Y.-M.; Zhou, B. T.; Law, K. T. Spin-Orbit-Parity-Coupled Superconductivity in Topological Monolayer $WTe_2$. *Phys. Rev. Lett.* **2020**, *125* (10), 107001.

(7) Kezilebieke, S.; Huda, M. N.; Vaňo, V.; Aapro, M.; Ganguli, S. C.; Silveira, O. J.; Głodzik, S.; Foster, A. S.; Ojanen, T.; Liljeroth, P. Topological Superconductivity in a van der Waals Heterostructure. *Nature* **2020**, *588* (7838), 424-428.

(8) Sasaki, S.; Kriener, M.; Segawa, K.; Yada, K.; Tanaka, Y.; Sato, M.; Ando, Y. Topological Superconductivity in $Cu_xBi_2Se_3$. *Phys. Rev. Lett.* **2011**, *107* (21), 217001.

(9) Xu, G.; Lian, B.; Tang, P.; Qi, X.-L.; Zhang, S.-C. Topological Superconductivity on the Surface of Fe-Based Superconductors. *Phys. Rev. Lett.* **2016**, *117* (4), 047001.

(10) Zhang, P.; Yaji, K.; Hashimoto, T.; Ota, Y.; Kondo, T.; Okazaki, K.; Wang, Z.; Wen, J.; Gu, G. D.; Ding, H.; Shin, S. Observation of Topological Superconductivity on the Surface of an Iron-based superconductor. *Science* **2018**, *360* (6385), 182-186.

(11) Lu, J. M.; Zheliuk, O.; Leermakers, I.; Yuan, N. F. Q.; Zeitler, U.; Law, K. T.; Ye, J. T. Evidence for Two-dimensional Ising Superconductivity in Gated $MoS_2$. *Science* **2015**, *350* (6266), 1353-1357.

(12) Lu, J.; Zheliuk, O.; Chen, Q.; Leermakers, I.; Hussey, N. E.; Zeitler, U.; Ye, J. Full Superconducting Dome of Strong Ising Protection in Gated Monolayer $WS_2$. *Proc. Natl. Acad. Sci. U.S.A.* **2018**, *115* (14), 3551-3556.

(13) Xi, X.; Wang, Z.; Zhao, W.; Park, J.-H.; Law, K. T.; Berger, H.; Forró, L.; Shan, J.; Mak, K. F. Ising Pairing in Superconducting $NbSe_2$ Atomic Layers. *Nat. Phys.* **2015**, *12* (2), 139-143.

(14) Fu, L.; Kane, C. L. Superconducting Proximity Effect and Majorana Fermions at the Surface of a Topological Insulator. *Phys. Rev. Lett.* **2008**, *100* (9), 096407.

(15) Nayak, C.; Simon, S. H.; Stern, A.; Freedman, M.; Das Sarma, S. Non-Abelian Anyons and Topological Quantum Computation. *Rev. Mod. Phys.* **2008**, *80* (3), 1083-1159.

(16) Flensberg, K.; von Oppen, F.; Stern, A. Engineered Platforms for Topological Superconductivity and Majorana Zero Modes. *Nat. Rev. Mater.* **2021**, *6* (10), 944-958.

(17) Fang, Y.; Pan, J.; Zhang, D.; Wang, D.; Hirose, H. T.; Terashima, T.; Uji, S.; Yuan, Y.; Li, W.; Tian, Z.; Xue, J.; Ma, Y.; Zhao, W.; Xue, Q.; Mu, G.; Zhang, H.; Huang, F. Discovery of Superconductivity in 2M $WS_2$ with Possible Topological Surface States. *Adv. Mater.* **2019**, *0* (0), 1901942.

(18) Che, X.; Deng, Y.; Fang, Y.; Pan, J.; Yu, Y.; Huang, F. Gate-Tunable Electrical Transport in Thin 2M-$WS_2$ Flakes. *Adv. Electron. Mater.* **2019**, *5* (10), 1900462.

(19) Guguchia, Z.; Gawryluk, D. J.; Brzezinska, M.; Tsirkin, S. S.; Khasanov, R.; Pomjakushina, E.; von Rohr, F. O.; Verezhak, J. A. T.; Hasan, M. Z.; Neupert, T.; Luetkens, H.; Amato, A. Nodeless

Superconductivity and its Evolution with Pressure in the Layered Dirac Semimetal 2M-WS$_2$. *npj Quantum Mater.* **2019**, *4* (1), 50.

(20) Yuan, Y.; Pan, J.; Wang, X.; Fang, Y.; Song, C.; Wang, L.; He, K.; Ma, X.; Zhang, H.; Huang, F.; Li, W.; Xue, Q.-K. Evidence of Anisotropic Majorana Bound States in 2M-WS$_2$. *Nat. Phys.* **2019**, *15* (10), 1046-1051.

(21) Wang, L. S.; Fang, Y. Q.; Huang, Y. Y.; Cheng, E. J.; Ni, J. M.; Pan, B. L.; Xu, Y.; Huang, F. Q.; Li, S. Y. Nodeless Superconducting Gap in the Topological Superconductor Candidate 2M WS$_2$. *Phys. Rev. B* **2020**, *102* (2), 024523.

(22) Joseph, N. B.; Narayan, A. Topological Properties of Bulk and Bilayer 2M WS$_2$: a First-principles Study. *J. Phys.: Condens. Matter* **2021**, *33* (46), 465001.

(23) Li, Y. W.; Zheng, H. J.; Fang, Y. Q.; Zhang, D. Q.; Chen, Y. J.; Chen, C.; Liang, A. J.; Shi, W. J.; Pei, D.; Xu, L. X.; Liu, S.; Pan, J.; Lu, D. H.; Hashimoto, M.; Barinov, A.; Jung, S. W.; Cacho, C.; Wang, M. X.; He, Y.; Fu, L.; Zhang, H. J.; Huang, F. Q.; Yang, L. X.; Liu, Z. K.; Chen, Y. L. Observation of Topological Superconductivity in a Stoichiometric Transition Metal Dichalcogenide 2M-WS$_2$. *Nat. Commun.* **2021**, *12* (1), 2874.

(24) Lian, C.-S.; Si, C.; Duan, W. Anisotropic Full-Gap Superconductivity in 2M-WS2 Topological Metal with Intrinsic Proximity Effect. *Nano Lett.* **2021**, *21* (1), 709-715.

(25) Cho, S.; Huh, S.; Fang, Y.; Hua, C.; Bai, H.; Jiang, Z.; Liu, Z.; Liu, J.; Chen, Z.; Fukushima, Y.; Harasawa, A.; Kawaguchi, K.; Shin, S.; Kondo, T.; Lu, Y.; Mu, G.; Huang, F.; Shen, D. Direct Observation of the Topological Surface State in the Topological Superconductor 2M-WS$_2$. *Nano Lett.* **2022**, *22* (22), 8827-8834.

(26) Zhang, E.; Xie, Y.-M.; Fang, Y.; Zhang, J.; Xu, X.; Zou, Y.-C.; Leng, P.; Gao, X.-J.; Zhang, Y.; Ai, L.; Zhang, Y.; Jia, Z.; Liu, S.; Yan, J.; Zhao, W.; Haigh, S. J.; Kou, X.; Yang, J.; Huang, F.; Law, K. T.; Xiu, F.; Dong, S. Spin–orbit–parity Coupled Superconductivity in Atomically Thin 2M-WS$_2$. *Nat. Phys.* **2023**, *19* (1), 106-113.

(27) Gautam, S.; McBride, J.; Scougale, W. R.; Samarawickrama, P. I.; De Camargo Branco, D.; Yang, P.; Fu, Z.; Wang, W.; Tang, J.; Cheng, G. J.; Ackerman, J.; Chien, T.; Leonard, B. M.; Tian, J. Controllable Superconducting to Semiconducting Phase Transition in Topological Superconductor 2M-WS$_2$. *2D Mater.* **2024**, *11* (1), 015018.

(28) Samarawickrama, P.; Dulal, R.; Fu, Z.; Erugu, U.; Wang, W.; Ackerman, J.; Leonard, B.; Tang, J.; Chien, T.; Tian, J. Two-Dimensional 2M-WS$_2$ Nanolayers for Superconductivity. *ACS Omega* **2021**, *6* (4), 2966-2972.

(29) Sharma, C. H.; Surendran, A. P.; Varma, S. S.; Thalakulam, M. 2D Superconductivity and Vortex Dynamics in 1T-MoS$_2$. *Commun. Phys.* **2018**, *1* (1), 90.

(30) Itahashi, Y. M.; Saito, Y.; Ideue, T.; Nojima, T.; Iwasa, Y. Quantum and Classical Ratchet Motions of Vortices in a Two-dimensional Trigonal Superconductor. *Phys. Rev. Res.* **2020**, *2* (2), 023127.

(31) Yazdani, A.; Kapitulnik, A. Superconducting-Insulating Transition in Two-Dimensional *a*-MoGe Thin Films. *Phys. Rev. Lett.* **1995**, *74* (15), 3037-3040.

(32) Tsen, A. W.; Hunt, B.; Kim, Y. D.; Yuan, Z. J.; Jia, S.; Cava, R. J.; Hone, J.; Kim, P.; Dean, C. R.; Pasupathy, A. N. Nature of the Quantum Metal in a Two-dimensional Crystalline Superconductor. *Nat. Phys.* **2016**, *12* (3), 208-212.

(33) Tinkham, M. *Introduction to Superconductivity*; Dover Publications, 2004.

(34) Chiang, C. C.; Lee, H. C.; Lin, S. C.; Qu, D.; Chu, M. W.; Chen, C. D.; Chien, C. L.; Huang, S. Y. Unequivocal Identification of Spin-Triplet and Spin-Singlet Superconductors with Upper Critical Field and Flux Quantization. *Phys. Rev. Lett.* **2023**, *131* (23), 236003.

(35) Llovo, I. F.; Carballeira, C.; Sóñora, D.; Pereiro, A.; Ponte, J. J.; Salem-Sugui, S.; Sefat, A. S.; Mosqueira, J. Multiband Effects on the Upper Critical Field Angular Dependence of 122-family Iron Pnictide Superconductors. *Sci. Rep.* **2021**, *11* (1), 11526.